\documentclass[12pt]{article}
\usepackage[left=1in,right=1in,top=1in,bottom=1in]{geometry}
\usepackage{graphicx}
\usepackage{float}
\usepackage{amsfonts}
\usepackage{amssymb}
\usepackage{amsmath}
\usepackage{relsize}

\def\gtwid{\mathrel{\raise.3ex\hbox{$>$\kern-.75em\lower1ex\hbox{$\sim$}}}}
\def\ltwid{\mathrel{\raise.3ex\hbox{$<$\kern-.75em\lower1ex\hbox{$\sim$}}}}
\def\square{\kern1pt\vbox{\hrule height 1.2pt\hbox{\vrule width 1.2pt\hskip 3pt
   \vbox{\vskip 6pt}\hskip 3pt\vrule width 0.6pt}\hrule height 0.6pt}\kern1pt}

\begin{document}

\begin{titlepage}

\begin{flushright}
UFIFT-QG-23-01
\end{flushright}

\vskip 4cm

\begin{center}
{\bf Perturbative Quantum Gravity Induced Scalar Coupling to Electromagnetism}
\end{center}

\vskip 1cm

\begin{center}
S. Katuwal$^{*}$ and R. P. Woodard$^{\dagger}$
\end{center}

\begin{center}
\it{Department of Physics, University of Florida, \\
Gainesville, FL 32611, UNITED STATES}
\end{center}

\vspace{1cm}

\begin{center}
ABSTRACT
\end{center}
Physicists working on atom interferometers are interested in scalar couplings to
electromagnetism of dimensions 5 and 6 which might be induced by quantum gravity. 
There is a widespread belief that such couplings can only be induced by conjectured 
non-perturbative effects, resulting in unknown coupling strengths. In this letter
we exhibit a completely perturbative mechanism through which quantum gravity 
induces dimension six couplings with precisely calculable coefficients.

\begin{flushleft}
\end{flushleft}

\vspace{6cm}

\begin{flushleft}
$^{*}$ e-mail: sanjib.katuwal@ufl.edu \\
$^{\dagger}$ e-mail: woodard@phys.ufl.edu
\end{flushleft}

\end{titlepage}

\section{Introduction}

There has been much recent interest in searching for exotic processes which
might be induced by quantum gravity \cite{Piscicchia:2022xra,Piscicchia:2022eod}.
In particular, it has been suggested \cite{Calmet:2022bin} that quantum gravitationally
induced scalar couplings to electromagnetism might be detected by planned atom 
interferometers such as MAGIS \cite{Abe:2021magis}, AION \cite{Badurina:2019hst,
Badurina:2021rgt} and AEDGE \cite{AEDGE:2019nxb}. Conventional wisdom has it that
perturbative quantum gravity can at best generate couplings of dimension eight, and
that couplings of dimensions 5 and 6 could only be induced, with unknown coefficients,
by nonperturbative effects \cite{Calmet:2019frv} such as gravitational instantons 
\cite{Perry:1978fd,Chen:2021jcb} and wormholes \cite{Gilbert:1989nq}.

In this paper we point out that there is a completely perturbative mechanism 
through which quantum gravity induces a dimension six coupling of a massive
scalar with a precisely calculable coefficient. The mechanism is simple: assuming 
that the scalar is constant in space and time, and that the potential energy 
from its mass dominates the stress-energy, the background geometry will be de 
Sitter with a Hubble parameter which depends in a precise way on the scalar. 
Unlike the graviton propagator in flat space, the coincidence limit of a
graviton propagator on de Sitter background goes like the square of the 
Hubble parameter in any gauge \cite{Tsamis:1992xa,Woodard:2004ut,Miao:2011fc,
Mora:2012zi,Glavan:2019msf}. Hence integrating out pairs of graviton fields 
from the Heisenberg operator Maxwell equation (the Hartree approximation) 
induces couplings of the desired form with precisely computable coefficients.

This short letter contains only three sections, of which this Introduction is 
the first. We present the calculation in Section \ref{sec:calculation}. Our 
conclusions comprise Section \ref{sec:conclusion}.

\section{Calculation}\label{sec:calculation}

Consider a massive, uncharged scalar field which is coupled to electromagnetism
and gravity,
\begin{equation}
\mathcal{L} = -\frac{1}{2}\partial_{\mu} \varphi \partial_{\nu} \varphi 
g^{\mu\nu} \sqrt{-g} - \frac{1}{2} m^2 \varphi^2 \sqrt{-g} -\frac{1}{4}
F_{\mu\nu} F_{\rho\sigma} g^{\mu\rho} g^{\nu\sigma} \sqrt{-g} -
\frac{R \sqrt{-g}}{16\pi G} \; . \label{eq:lagrangian}
\end{equation}
The corresponding Einstein equation is 
\begin{equation}
G_{\mu\nu} = 8\pi G \left[\partial_{\mu} \varphi \partial_{\nu} \varphi - 
\frac{1}{2} g_{\mu\nu} g^{\rho\sigma} \partial_{\rho} \varphi \partial_{\sigma}
\varphi - \frac{1}{2} m^2 \varphi^2 g_{\mu\nu} + F_{\mu\rho} F_{\nu\sigma}
g^{\rho\sigma} - \frac{1}{4} g_{\mu\nu} g^{\alpha\beta} g^{\rho\sigma}
F_{\alpha\rho} F_{\beta\sigma} \right] . \label{eq:einstein}
\end{equation}
As discussed in the introduction, we assume that $\varphi=\varphi_0$ is a 
constant and also set $A_\mu=0$ to get,
\begin{equation}
G_{\mu\nu} = 8 \pi G \times -\frac{1}{2} m^2 \varphi_0^2 g_{\mu\nu} \; .
\label{eq:einsteinReduced}
\end{equation}
The unique, maximally symmetric solution is de Sitter with Hubble constant,
\begin{equation}
H = \sqrt{\frac{4\pi G}{3} \times m^2 \varphi_0^2} \; . \label{eq:deSitterH}
\end{equation}
where, $H$ is the Hubble constant. This shows that a constant scalar triggers
a phase of de Sitter inflation. Because de Sitter is conformally flat, there is 
no classical effect on electromagnetism in conformal coordinates. However, we
will see that the breaking of conformal invariance by gravity does induce a
quantum effect.

Consider the Maxwell equation in a general metric $g_{\mu\nu}$,
\begin{equation}
\partial_{\nu} \Bigl[\sqrt{-g} \, g^{\mu\rho} g^{\nu\sigma} F_{\sigma\rho}
\Bigr] = J^{\mu} \; , \label{eq:maxwell}
\end{equation}
where $F_{\mu\nu} \equiv \partial_{\mu} A_{\nu} - \partial_{\nu} A_{\mu}$ is the 
field strength tensor and $J^{\mu}$ is the current density. We write the quantum
metric $g_{\mu\nu}$ in terms of the Minkowski metric $\eta_{\mu\nu}$,
\begin{equation}
g_{\mu\nu} \equiv a^2 \tilde{g}_{\mu\nu} = a^2 \Bigl[\eta_{\mu\nu} + \kappa 
h_{\mu\nu} \Bigr] \; , \label{eq:gmunuPerturb}
\end{equation}
where $\kappa^2 \equiv 16 \pi G$ is the loop counting parameter, $a(\eta)
\equiv -1/H\eta$ is the scale factor at conformal time $\eta$ and $h_{\mu\nu}$ 
is the graviton field. Graviton indices are raised and lowered with the Minkowski
metric: $h^{\mu}_{~\nu} \equiv \eta^{\mu\rho} h_{\rho\nu}$, $h^{\mu\nu} \equiv
\eta^{\mu\rho} \eta^{\nu\sigma} h_{\rho\sigma}$. The inverse and determinant of 
the conformally transformed metric are,
\begin{eqnarray}
\tilde{g}^{\alpha\beta} & = & \eta^{\alpha\beta} - \kappa h^{\alpha\beta} + 
\kappa^2 h^{\alpha\rho} h^{\beta}_{~\rho} - \cdots \; , \label{eq:gInverse} \\
\sqrt{-\tilde{g}} & = & 1 + \frac{1}{2} \kappa h + \kappa^2 \left(\frac{1}{8}
h^2 - \frac{1}{4} h^{\rho\sigma} h_{\rho\sigma}\right) + \cdots \label{eq:detg}
\end{eqnarray}
Here $h$ is the trace of the graviton field $h \equiv \eta^{\mu\nu} h_{\mu\nu}$.

To facilitate dimensional regularization we formulate the theory in $D$ spacetime
dimensions. The term inside the square bracket of equation (\ref{eq:maxwell}) can
be expressed in terms of the conformally transformed metric as $\sqrt{-g} \, 
g^{\mu\rho} g^{\nu\sigma} F_{\sigma\rho} \equiv a^{D-4} \sqrt{-\tilde{g}} \, 
\tilde{g}^{\mu\rho} \tilde{g}^{\nu\sigma} F_{\sigma\rho}$. The terms involving 
$\tilde{g}_{\mu\nu}$ can be expanded as,
\begin{eqnarray}
\lefteqn{\tilde{g}^{\mu\rho} \tilde{g}^{\nu\sigma} \sqrt{-\tilde{g}} =
\eta^{\mu\rho} \eta^{\nu\sigma} + \kappa \Bigl[\frac{1}{2} h \eta^{\mu\rho}
\eta^{\nu\sigma} - h^{\mu\rho} \eta^{\nu\sigma} - \eta^{\mu\rho} h^{\nu\sigma}
\Bigr] + \kappa^2 \Biggl[\Bigl(\frac{1}{8} h^2 \!-\! \frac{1}{4} 
h^{\alpha\beta} h_{\alpha\beta} \Bigr) \eta^{\mu\rho} \eta^{\nu\sigma} }
\nonumber \\
& & \hspace{3cm} - \frac{1}{2} h h^{\mu\rho} \eta^{\nu\sigma} \!-\! \frac{1}{2} 
h\eta^{\mu\rho} h^{\nu\sigma} \!+\! h^{\mu\alpha} h^{\rho}_{~\alpha} 
\eta^{\nu\sigma} \!+\! \eta^{\mu\rho} h^{\nu\alpha} h^{\sigma}_{~\alpha} \!+\!
h^{\mu\rho} h^{\nu\sigma} \Biggr] \!+\! \mathcal{O}(\kappa^3) \; . \qquad 
\label{eq:expansion}
\end{eqnarray}
Using the Hartree approximation \cite{Hartree:1928,Mora:2013ypa}, we can replace the 
terms proportional to $\kappa$ by zero and the terms proportional to $\kappa^2$ by 
the coincidence limit of the graviton propagator $i[\mbox{}_{\mu\nu} 
\Delta_{\rho\sigma}](x;x')$,
\begin{eqnarray}
\lefteqn{ \tilde{g}^{\mu\rho} \tilde{g}^{\nu\sigma} \sqrt{-\tilde{g}} \longrightarrow
\eta^{\mu\rho} \eta^{\nu\sigma} + \kappa^2 \Biggl[ \Bigl(\frac{1}{8} 
i\Bigl[\mbox{}^{\alpha}_{~\alpha} \Delta^{\beta}_{~\beta}\Bigr] - \frac{1}{4}
i \Bigl[\mbox{}^{\alpha\beta} \Delta_{\alpha\beta}\Bigr] \Bigr) \eta^{\mu\rho}
\eta^{\nu\sigma} - \frac{1}{2} i \Bigl[\mbox{}^{\alpha}_{~\alpha} \Delta^{\mu\rho}
\Bigr] \eta^{\nu\sigma} } \nonumber \\
& & \hspace{2.5cm} - \frac{1}{2} i \Bigl[\mbox{}^{\alpha}_{~\alpha} \Delta^{\nu\sigma} 
\Bigr] \eta^{\mu\rho} + i \Bigl[\mbox{}^{\mu\alpha} \Delta^{\rho}_{~\alpha}\Bigr] 
\eta^{\nu\sigma} + \eta^{\mu\rho} i \Bigl[\mbox{}^{\nu\alpha} \Delta^{\sigma}_{~\alpha}
\Bigr] + i \Bigl[\mbox{}^{\mu\rho} \Delta^{\nu\sigma}\Bigr] \Biggr] + 
\mathcal{O}(\kappa^4) \; . \qquad \label{eq:kappaSqr}
\end{eqnarray}

The graviton is of course gauge dependent but its coincidence limit on de Sitter is
proportional to $H^{D-2}$ in all gauges \cite{Tsamis:1992xa,Woodard:2004ut,Miao:2011fc,
Mora:2012zi,Glavan:2019msf}. In the simplest gauge \cite{Tsamis:1992xa,Woodard:2004ut}
it consists of a sum of three constant tensor factors, each multiplied by a different
scalar propagator,
\begin{equation}
i\Bigl[\mbox{}_{\mu\nu} \Delta_{\alpha\beta}\Bigr](x;x') = \sum_{I=A,B,C}
\Bigl[\mbox{}_{\mu\nu} T^I_{\alpha\beta}\Bigr] \times i\Delta_I(x;x') \; .
\label{eq:gPropagator}
\end{equation}
The constant tensor factors are,
\begin{eqnarray}
\Bigl[\mbox{}_{\mu\nu} T^A_{\alpha\beta}\Bigr] & = & 2 \, \bar{\eta}_{\mu (\alpha} 
\bar{\eta}_{\beta )\nu} - \frac{2}{D-3} \, \bar{\eta}_{\mu\nu} \bar{\eta}_{\alpha\beta}
\quad , \quad \Bigl[\mbox{}_{\mu\nu} T^B_{\alpha\beta}\Bigr] = - 4 
\delta^{0}_{~(\mu} \bar{\eta}_{\nu)(\alpha} \delta^{0}_{~\beta)} \; , \\
\Bigl[\mbox{}_{\mu\nu} T^C_{\alpha\beta}\Bigr] & = & \frac{2}{(D-2)(D-3)} \Bigl[
\bar{\eta}_{\mu\nu} + (D-3) \delta^{0}_{~\mu} \delta^{0}_{~\nu} \Bigr] \Bigl[
\bar{\eta}_{\alpha\beta} + (D-3) \delta^{0}_{~\alpha} \delta^{0}_{~\beta}\Bigr] \; .
\qquad \label{eq:gPropTensorFactors}
\end{eqnarray}
where parenthesized indices are symmetrized and $\bar{\eta}_{\mu\nu} \equiv 
\eta_{\mu\nu} + \delta^{0}_{~\mu} \delta^{0}_{~\nu}$ is the purely spatial part of 
the Minkowski metric. The three scalar propagators correspond to masses $M_A^2 = 0$,
$M_B^2 = (D-2) H^2$ and $M_C^2 = 2 (D-3) H^2$. They obey the propagator equations,
\begin{equation}
\mathcal{D}_I i\Delta_I(x;x') = i \delta^D(x-x') \; , \label{eq:scalarEquation}
\end{equation}
where the various kinetic operators are,
\begin{equation}
\mathcal{D}_I \equiv \partial_{\mu} \Bigl( a^{D-2} \eta^{\mu\nu} \partial_{\nu} \Bigr)
- M_I^2 a^{D} \; . \label{eq:opeerators}
\end{equation}
The coincidence limits of the three scalar propagators are \cite{Glavan:2019msf},
\begin{eqnarray}
i\Delta_A(x;x) & = & k \Bigl[-\pi \cot\Bigl(\frac{D\pi}{2}\Bigr) + 2\ln(a) \Bigr]
\; , \label{eq:scalarA} \\
i \Delta_B(x;x)& = & -\frac{k}{D-2}\Bigl\vert_{D \rightarrow 4} =
-\frac{H^2}{16\pi^2} \; , \label{eq:scalarB} \\
i\Delta_C(x;x) & = & \frac{k}{(D-3)(D-2)}\Bigl\vert_{D \rightarrow 4} =
+\frac{H^2}{16\pi^2} \; , \label{eq:scalarC}
\end{eqnarray}
where the constant $k$ is, 
\begin{equation}
k \equiv \frac{H^{D-2}}{(4\pi)^{\frac{D}2}} \frac{\Gamma{(D-1)}}{\Gamma(\frac{D}2)}
\Bigl\vert_{D \rightarrow 4} = \frac{H^2}{8\pi^2} \; . \label{eq:kDefine}
\end{equation}

By employing the relations (\ref{eq:gPropagator}-\ref{eq:gPropTensorFactors}) 
and (\ref{eq:scalarA}-\ref{eq:scalarC}) which define the coincident graviton
propagator in expression (\ref{eq:kappaSqr}), and then substituting into the 
left hand side of Maxwell's equation (\ref{eq:maxwell}), we obtain the order
$\kappa^2$ correction,
\begin{eqnarray}
\lefteqn{\partial_{\nu} \Bigl[\sqrt{-g} \, g^{\mu\rho} g^{\nu\sigma} 
F_{\sigma\rho} \Bigr]_{\kappa^2} = \frac{\kappa^2 H^2}{4\pi^2} \Biggl[ \tfrac{1}{4}
\Bigl(\partial_{\nu} F^{\nu}_{~0} \delta^{\mu}_{~0} + \partial_0 F_{0}^{~\mu}
\Bigr) + \partial_{\nu} \Bigl\{ (F_{0}^{~\mu} \delta_{0}^{~\nu} \!+\!
F^{\nu}_{~0} \delta_{0}^{~\mu} \!-\! 2 F^{\mu\nu} ) \ln(a) \Bigr\} \Biggr]}
\nonumber \\
& & \hspace{0.5cm} -\kappa^2 k \pi \cot(\tfrac{D \pi}{2}) \partial_{\nu} 
\Biggl[ a^{D-4} \Bigl\{ (\tfrac{D^2 - 4 D + 1}{D - 3})
\Bigl( F_{0}^{~\mu} \delta_{0}^{~\nu} \!+\! F^{\nu}_{~0} \delta_{0}^{~\mu} 
\Bigr) + (\tfrac{D^3 - 12 D^2 + 31 D - 4}{4 (D - 3)}) F^{\mu\nu} \Bigr\} \Biggr] 
. \qquad \label{eq:maxwellFinal}
\end{eqnarray}
Renormalization is facilitated by expressing the divergent part in terms of 
the purely spatial components $\bar{F}^{\mu\nu} \equiv \bar{\eta}^{\mu\rho} 
\bar{\eta}^{\nu\sigma} F_{\rho\sigma}$ of the field strength tensor,
\begin{eqnarray}
\lefteqn{\partial_{\nu} \Bigl[\sqrt{-g} \, g^{\mu\rho} g^{\nu\sigma} 
F_{\sigma\rho} \Bigr]_{\kappa^2} = \frac{\kappa^2 H^2}{4 \pi^2} \Biggl[
\tfrac{1}{4} \partial_{\nu} \Bigl\{ F^{\mu\nu} - \bar{F}^{\mu\nu} \Bigr\} - 
\partial_{\nu} \Bigl\{ (F^{\mu\nu} \!+\! \bar{F}^{\mu\nu} ) \ln(a) \Bigr\} 
\Biggr]} \nonumber \\
& & \hspace{4.3cm} + \kappa^2 k \pi \cot(\tfrac{D \pi}{2}) \partial_{\nu} 
\Biggl[ a^{D-4} \Bigl\{ -\tfrac{D (D - 5)}{4} F^{\mu\nu} + 
(\tfrac{D^2 - 4 D + 1}{D - 3}) \bar{F}^{\mu\nu} \Bigr\} \Biggr] . \qquad
\label{eq:maxwellFinal1}
\end{eqnarray}
The cotangent is divergent as $D \rightarrow 4$,
\begin{equation}
\pi \cot(\tfrac{D \pi}{2}) = \frac{2}{D - 4} + O(D \!-\! 4) \; .
\end{equation}
The divergences on the final line of (\ref{eq:maxwellFinal1}) can be eliminated
by the counterterm,
\begin{equation}
\Delta\mathcal{L} = \tfrac{\kappa^2}{4} (\tfrac{\mu}{H})^{D-4} k \pi 
\cot(\tfrac{D \pi}{2}) \Bigl\{ -\tfrac{D (D - 5)}{4} F_{\alpha\beta}
F_{\rho\sigma} + (\tfrac{D^2 - 4 D + 1}{D - 3}) \bar{F}_{\alpha\beta} 
\bar{F}_{\rho\sigma} \Bigr\} \frac{R g^{\alpha\rho} g^{\beta\sigma} \sqrt{-g}}{
(D \!-\! 1) D H^2} \; . \label{eq:counterterm}
\end{equation}
Note the factor of $(\mu/H)^{D-4}$ required to cancel the $D$-dependence
in the factor of $H^{D-2}$ evident in expression (\ref{eq:kDefine}) for the 
constant $k$. Note also that the need for a noninvariant counterterm arises
from the avoidable breaking of de Sitter invariance in the simplest gauge
\cite{Tsamis:1992xa,Woodard:2004ut} and from the unavoidable time-ordering
of interactions \cite{Glavan:2015ura}.

Combining the variation of the counterterm (\ref{eq:counterterm}) with the
primitive contribution (\ref{eq:maxwellFinal1}), and then taking the unregulated
limit gives,
\begin{equation}
\partial_{\nu} \Bigl[\sqrt{-g} \, g^{\mu\rho} g^{\nu\sigma} 
F_{\sigma\rho} \Bigr]_{\kappa^2} = \frac{\kappa^2 H^2}{4 \pi^2} \Biggl[
\tfrac{1}{4} \partial_{\nu} \Bigl\{ F^{\mu\nu} \!-\! \bar{F}^{\mu\nu} \Bigr\} - 
\partial_{\nu} \Bigl\{ (F^{\mu\nu} \!+\! \bar{F}^{\mu\nu} ) 
\ln\Bigl(\frac{\mu a}{H}\Bigr) \Bigr\} \Biggr] . \label{eq:maxwellFinal2}
\end{equation}
Note the $\mu$-dependence against the scale factor $a(\eta)$ in the logarithm
at the end. This is a vestige of renormalization. Substituting for the Hubble 
constant from expression (\ref{eq:deSitterH}), and recalling that the 
loop-counting parameter is $\kappa^2 = 16 \pi G$, results in the final 
dimension six coupling to Maxwell's equation,
\begin{equation}
\partial_{\nu} \Bigl[\sqrt{-g} \, g^{\mu\rho} g^{\nu\sigma} 
F_{\sigma\rho} \Bigr]_{\kappa^2} = \tfrac{16}{3} G^2 m^2 \phi^2 \Biggl[
\tfrac{1}{4} \partial_{\nu} \Bigl\{ F^{\mu\nu} \!-\! \bar{F}^{\mu\nu} \Bigr\} - 
\partial_{\nu} \Bigl\{ (F^{\mu\nu} \!+\! \bar{F}^{\mu\nu} ) 
\ln\Bigl(\frac{\sqrt{3} \, \mu a}{\sqrt{4\pi G} \, m \phi}\Bigr) \Bigr\} 
\Biggr] . \label{eq:maxwellFinal3}
\end{equation}
Note that the scalar could have as easily been placed inside the $\partial_{\nu}$
--- as it would have been in varying the counterterm (\ref{eq:counterterm}) ---
because the computation was made assuming $\phi$ and the induced $H$ were 
constant. Although a finite renormalization could have eliminated the term 
proportional to $F^{\mu\nu} - \bar{F}^{\mu\nu}$ in (\ref{eq:maxwellFinal3}), the 
logarithm of $\phi$ multiplying the other term is a genuine prediction of the 
theory, with a specific coefficient which we have just computed.

\section{Conclusion}\label{sec:conclusion}

The usual way a constant scalar background engenders quantum corrections
is by giving some field a mass so that the coincidence limit of that field's
propagator depends on the scalar. That cannot happen in perturbative quantum
gravity because the graviton remains massless to all orders. However, 
constant scalars can also contribute by changing a field strength 
\cite{Miao:2021gic}. In our case, a constant scalar background changes the 
cosmological constant, and the graviton propagator in de Sitter background 
depends upon this cosmological constant \cite{Tsamis:1992xa,Woodard:2004ut,
Miao:2011fc,Mora:2012zi,Glavan:2019msf}. We have exploited this mechanism to 
compute the dimension six coupling (\ref{eq:maxwellFinal3}) to 
electromagnetism. Similar results could be obtained for couplings to any
other low energy field.

Our result comes with several important caveats, both theoretical and 
phenomenological. On the theoretical side, our computation depended on the 
scalar being constant throughout spacetime. Although this is not a reasonable 
assumption, setting the scalar to be constant is the correct way to compute 
nonderivative couplings, which should remain valid in the resulting low energy 
effective field theory, even when the assumption of constancy is relaxed. We 
have also assumed that the scalar potential energy dominates the total stress 
energy, which is not reasonable for a weak scalar. However, the coupling must 
still be present in a realistic cosmological background because it must be 
there in the large field limit. Finally, although the graviton propagator is 
gauge dependent, dimensional analysis requires its coincidence limit on de 
Sitter background to go like $H^{D-2}$, a fact which is confirmed in all 
known gauges \cite{Tsamis:1992xa,Woodard:2004ut,Miao:2011fc,Mora:2012zi,
Glavan:2019msf}. A recently developed formalism \cite{Miao:2017feh,
Katuwal:2020rkv}, based on the S-matrix, can be used to remove gauge 
dependence from the effective field equations.

It should be possible to verify our beliefs by employing the same techniques 
which were recently used to compute cosmological Coleman-Weinberg potentials 
\cite{Kyriazis:2019xgj,Sivasankaran:2020dzp,Katuwal:2021kry,
Katuwal:2022szw}. In this regard, consider the graviton propagator for an
evolving cosmological geometry with Hubble parameter $H(t)$ and first slow roll 
parameter $\epsilon(t)$,
\begin{equation}
ds^2 = -dt^2 + a^2(t) d\vec{x} \!\cdot\! d\vec{x} \quad \Longrightarrow \quad
H(t) \equiv \frac{\dot{a}}{a} \;\; , \;\; \epsilon(t) \equiv -\frac{\dot{H}}{H^2}
\; . \label{geometry}
\end{equation}
This geometry is supported by the energy density $\rho$ and pressure $p$ of 
matter through the Friedmann equations,
\begin{eqnarray}
(D\!-\!1) H^2 & \!\!\! = \!\!\! & 8\pi G \rho \; , \label{Feqn1} \\
-(D\!-\!1) H^2 - 2 \dot{H} & \!\!\! = \!\!\! & 8 \pi G p \; . \label{Feqn2}
\end{eqnarray}
Our scalar adds to the energy density and pressure, but it need not dominate
them, nor need it be constant. Although the graviton propagator is not known for
general $\epsilon(t)$, it is known for the important special case of constant
$\epsilon$ \cite{Iliopoulos:1998wq}, which includes well-known epochs of 
cosmological evolution such as radiation domination ($\epsilon = 2$), matter 
domination ($\epsilon = \frac32$), and vacuum energy domination ($\epsilon = 0$).
The full propagator can be found in equation (3.38) of  \cite{Iliopoulos:1998wq},
but the part involving dynamical gravitons is a constant tensor factor times the 
propagator of the massless, minimally coupled scalar. The coincidence limit of 
this propagator is given by equation (33) of \cite{Janssen:2008px},
\begin{equation}
i\Delta(x;x) = -\frac{[(1\!-\!\epsilon) H(t)]^{D-2}}{(4\pi)^{\frac{D}{2}}}
\frac{\Gamma(\frac{D}{2} \!-\! 1) \Gamma(2 \!-\! \frac{D}{2})}{\Gamma(\frac{D}2)}
\frac{\Gamma(\frac{D-1}{2} \!+\! \nu) \Gamma(\frac{D-1}{2} \!-\! \nu)}{
\Gamma(\frac12 \!+\! \nu) \Gamma(\frac12 \!-\! \nu)} \qquad \nu \equiv
\frac{(D\!-\!1\!-\!\epsilon)}{2 (1 \!-\! \epsilon)} \; . \label{MMCS}
\end{equation}
Note the initial factor of $[H(t)]^{D-2}$, which can be written in terms of the 
total energy density $\rho$ using the first Friedmann equation (\ref{Feqn1}). This
remains true even if the scalar is only one part of a much larger contribution, and
even if the scalar is not constant in time. We believe that a similar result 
pertains for a scalar which is not constant in space.

An important phenomenological caveat concerns the fact that our coupling 
(\ref{eq:maxwellFinal3}) is proportional to $G^2 m^2 \phi^2$, rather than $G \phi^2$.
Although both couplings correspond to contributions to the action of dimension six, 
the extra factor of $G m^2$ in our result (\ref{eq:maxwellFinal3}) renders it 
unobservably small for the optimal mass of $m \sim 10^{-24}~{\rm GeV}$ which is 
relevant to atom interferometers \cite{Calmet:2022bin}. So our work must be regarded 
as an illustration of how perturbative quantum gravity can induce couplings to low
energy fields rather than as a testable prediction.

\vskip .5cm 

\centerline{\bf Acknowledgements}

We are grateful for discussions with C. Curceanu. This work was 
supported by NSF grant  PHY-2207514 and by the Institute for 
Fundamental Theory at the University of Florida.


\begin{thebibliography}{99}

\bibitem{Piscicchia:2022xra}
K.~Piscicchia, A.~Addazi, A.~Marciano, M.~Bazzi, M.~Cargnelli, A.~Clozza, L.~De Paolis, R.~Del Grande, C.~Guaraldo and M.~A.~Iliescu, \textit{et al.}
Phys. Rev. Lett. \textbf{129}, no.13, 131301 (2022)
doi:10.1103/PhysRevLett.129.131301
[arXiv:2209.00074 [hep-th]].

\bibitem{Piscicchia:2022eod}
K.~Piscicchia, A.~Addazi, A.~Marciano, M.~Bazzi, M.~Cargnelli, A.~Clozza, L.~De Paolis, R.~Del Grande, C.~Guaraldo and M.~A.~Iliescu, \textit{et al.}
[arXiv:2212.04669 [hep-th]].

\bibitem{Calmet:2022bin}
X.~Calmet and N.~Sherrill,
Universe \textbf{8}, no.2, 103 (2022)
doi:10.3390/universe8020103
[arXiv:2202.01458 [hep-ph]].

\bibitem{Abe:2021magis}
M. Abe, P. Adamson, \textit{et al}. 
Quantum Sci Technol. \textbf{6.4}(2021): 044003
[arXiv:2104.02835 [physics.atom-ph]].

\bibitem{Badurina:2019hst}
L.~Badurina, E.~Bentine, D.~Blas, K.~Bongs, D.~Bortoletto, T.~Bowcock, K.~Bridges, W.~Bowden, O.~Buchmueller and C.~Burrage, \textit{et al.}
JCAP \textbf{05}, 011 (2020)
doi:10.1088/1475-7516/2020/05/011
[arXiv:1911.11755 [astro-ph.CO]].

\bibitem{Badurina:2021rgt}
L.~Badurina, O.~Buchmueller, J.~Ellis, M.~Lewicki, C.~McCabe and V.~Vaskonen,
 dark matter,''
Phil. Trans. A. Math. Phys. Eng. Sci. \textbf{380}, no.2216, 20210060 (2021)
doi:10.1098/rsta.2021.0060
[arXiv:2108.02468 [gr-qc]].

\bibitem{AEDGE:2019nxb}
Y.~A.~El-Neaj \textit{et al.} [AEDGE],
EPJ Quant. Technol. \textbf{7}, 6 (2020)
doi:10.1140/epjqt/s40507-020-0080-0
[arXiv:1908.00802 [gr-qc]].

\bibitem{Calmet:2019frv}
X.~Calmet,
Phys. Lett. B \textbf{801}, 135152 (2020)
doi:10.1016/j.physletb.2019.135152
[arXiv:1912.04147 [hep-ph]].

\bibitem{Perry:1978fd}
M.~J.~Perry,
Phys. Rev. D \textbf{19}, 1720 (1979)
doi:10.1103/PhysRevD.19.1720

\bibitem{Chen:2021jcb}
Z.~Chen and A.~Kobakhidze,
Eur. Phys. J. C \textbf{82}, no.7, 596 (2022)
doi:10.1140/epjc/s10052-022-10542-3
[arXiv:2108.05549 [hep-ph]].

\bibitem{Gilbert:1989nq}
G.~Gilbert,
Nucl. Phys. B \textbf{328}, 159-170 (1989)
doi:10.1016/0550-3213(89)90097-7

\bibitem{Tsamis:1992xa}
N.~C.~Tsamis and R.~P.~Woodard,
Commun. Math. Phys. \textbf{162}, 217-248 (1994)
doi:10.1007/BF02102015

\bibitem{Woodard:2004ut}
R.~P.~Woodard,
[arXiv:gr-qc/0408002 [gr-qc]].

\bibitem{Miao:2011fc}
S.~P.~Miao, N.~C.~Tsamis and R.~P.~Woodard,
J. Math. Phys. \textbf{52}, 122301 (2011)
doi:10.1063/1.3664760
[arXiv:1106.0925 [gr-qc]].

\bibitem{Mora:2012zi}
P.~J.~Mora, N.~C.~Tsamis and R.~P.~Woodard,
J. Math. Phys. \textbf{53}, 122502 (2012)
doi:10.1063/1.4764882
[arXiv:1205.4468 [gr-qc]].

\bibitem{Glavan:2019msf}
D.~Glavan, S.~P.~Miao, T.~Prokopec and R.~P.~Woodard,
JHEP \textbf{10}, 096 (2019)
doi:10.1007/JHEP10(2019)096
[arXiv:1908.06064 [gr-qc]].

\bibitem{Hartree:1928}
D.~R.~Hartree,
Math. Proc. Camb. Philos. Soc. \textbf{24}, 111 (1928)
doi:10.1017/S0305004100011920.

\bibitem{Mora:2013ypa}
P.~J.~Mora, N.~C.~Tsamis and R.~P.~Woodard,
JCAP \textbf{10}, 018 (2013)
doi:10.1088/1475-7516/2013/10/018
[arXiv:1307.1422 [gr-qc]].

\bibitem{Glavan:2015ura}
D.~Glavan, S.~P.~Miao, T.~Prokopec and R.~P.~Woodard,
Class. Quant. Grav. \textbf{32}, no.19, 195014 (2015)
doi:10.1088/0264-9381/32/19/195014
[arXiv:1504.00894 [gr-qc]].

\bibitem{Miao:2021gic}
S.~P.~Miao, N.~C.~Tsamis and R.~P.~Woodard,
JHEP \textbf{03}, 069 (2022)
doi:10.1007/JHEP03(2022)069
[arXiv:2110.08715 [gr-qc]].

\bibitem{Miao:2017feh}
S.~P.~Miao, T.~Prokopec and R.~P.~Woodard,
Phys. Rev. D \textbf{96}, no.10, 104029 (2017)
doi:10.1103/PhysRevD.96.104029
[arXiv:1708.06239 [gr-qc]].

\bibitem{Katuwal:2020rkv}
S.~Katuwal and R.~P.~Woodard,
JHEP \textbf{21}, 029 (2020)
doi:10.1007/JHEP10(2021)029
[arXiv:2107.13341 [gr-qc]].

\bibitem{Kyriazis:2019xgj}
A.~Kyriazis, S.~P.~Miao, N.~C.~Tsamis and R.~P.~Woodard,
Phys. Rev. D \textbf{102}, no.2, 025024 (2020)
doi:10.1103/PhysRevD.102.025024
[arXiv:1908.03814 [gr-qc]].

\bibitem{Sivasankaran:2020dzp}
A.~Sivasankaran and R.~P.~Woodard,
Phys. Rev. D \textbf{103}, no.12, 125013 (2021)
doi:10.1103/PhysRevD.103.125013
[arXiv:2007.11567 [gr-qc]].

\bibitem{Katuwal:2021kry}
S.~Katuwal, S.~P.~Miao and R.~P.~Woodard,
Phys. Rev. D \textbf{103}, no.10, 105007 (2021)
doi:10.1103/PhysRevD.103.105007
[arXiv:2101.06760 [gr-qc]].

\bibitem{Katuwal:2022szw}
S.~Katuwal, S.~P.~Miao and R.~P.~Woodard,
JCAP \textbf{11}, 026 (2022)
doi:10.1088/1475-7516/2022/11/026
[arXiv:2208.11146 [hep-ph]].

\bibitem{Iliopoulos:1998wq}
J.~Iliopoulos, T.~N.~Tomaras, N.~C.~Tsamis and R.~P.~Woodard,
Nucl. Phys. B \textbf{534}, 419-446 (1998)
doi:10.1016/S0550-3213(98)00528-8
[arXiv:gr-qc/9801028 [gr-qc]].

\bibitem{Janssen:2008px}
T.~M.~Janssen, S.~P.~Miao, T.~Prokopec and R.~P.~Woodard,
Class. Quant. Grav. \textbf{25}, 245013 (2008)
doi:10.1088/0264-9381/25/24/245013
[arXiv:0808.2449 [gr-qc]].

\end{thebibliography}
\end{document}